\documentclass[graybox]{svmult}

\def\be{\begin{eqnarray}}
\def\ee{\end{eqnarray}}

\usepackage{mathptmx}       
\usepackage{helvet}         
\usepackage{courier}        
\usepackage{type1cm}        
\usepackage{makeidx}         
\usepackage{graphicx}        
\usepackage{multicol}        
\usepackage[bottom]{footmisc}

\makeindex             

\begin{document}

\title*{The QGP phase in relativistic heavy-ion collisions}

\author{E. L. Bratkovskaya, V. P. Konchakovski,
V. Voronyuk, V. D. Toneev, O.~Linnyk, and W. Cassing}
\institute{E. L. Bratkovskaya
\at Institute for Theoretical Physics, University of Frankfurt, Frankfurt, Germany;
\at Frankfurt Institute for Advanced Study, Frankfurt am Main, Germany\\
\email{Elena.Bratkovskaya@th.physik.uni-frankfurt.de}
\and  V. P. Konchakovski
\at Institute for Theoretical Physics, University of Giessen, Giessen, Germany
\and V. Voronyuk
\at Bogolyubov Institute for Theoretical Physics, Kiev, Ukraine;
\at  Joint Institute for Nuclear Research,  Dubna, Russia;
\at Frankfurt Institute for Advanced Study, Frankfurt am Main, Germany
\and V. D. Toneev
\at  Joint Institute for Nuclear Research,  Dubna, Russia;
\at Frankfurt Institute for Advanced Study, Frankfurt am Main, Germany
\and  O. Linnyk
\at Institute for Theoretical Physics, University of Giessen, Giessen, Germany
\and W. Cassing
\at Institute for Theoretical Physics, University of Giessen, Giessen, Germany
}
\authorrunning{E. L. Bratkovskaya et al.}

\maketitle

\abstract{
The dynamics of partons, hadrons and strings in relativistic
nucleus-nucleus collisions is analyzed within the novel
Parton-Hadron-String Dynamics (PHSD) transport approach, which is based
on a dynamical quasiparticle model for partons (DQPM) matched to
reproduce recent lattice-QCD results - including the partonic equation
of state - in thermodynamic equilibrium.  The transition from partonic
to hadronic degrees of freedom is described by covariant transition
rates for the fusion of quark-antiquark pairs or three quarks
(antiquarks), respectively, obeying flavor current-conservation, color
neutrality as well as energy-momentum conservation.
The PHSD approach is applied to nucleus-nucleus collisions from low SIS
to RHIC energies. The traces of partonic interactions are found in
particular in the elliptic flow of hadrons as well as in their
transverse mass spectra.  }


\section{Introduction}

The 'Big Bang' scenario implies that in the first micro-seconds of the
universe the entire state has emerged from a partonic system of
quarks, antiquarks and gluons -- a quark-gluon plasma (QGP) -- to
color neutral hadronic matter consisting of interacting hadronic
states (and resonances) in which the partonic degrees of freedom are
confined. The nature of confinement and the dynamics of this phase
transition has motivated a large community for several decades and is
still an outstanding question of todays physics. Early concepts of the
QGP were guided by the idea of a weakly interacting system of partons
which might be described by perturbative QCD (pQCD). However,
experimental observations at the Relativistic Heavy Ion Collider
(RHIC) indicated that the new medium created in ultrarelativistic
Au+Au collisions is interacting more strongly than hadronic matter and
consequently this concept had to be severely questioned. Moreover, in
line with theoretical studies in Refs.~\cite{Shuryak,Thoma,Andre} the
medium showed phenomena of an almost perfect liquid of
partons~\cite{STARS,Miklos3} as extracted from the strong radial
expansion and the scaling of elliptic flow $v_2(p_T)$ of mesons and
baryons with the number of constituent quarks and
antiquarks~\cite{STARS}.

The question about the properties of this (nonperturbative) QGP liquid
is discussed controversially in the literature and dynamical concepts
describing the formation of color neutral hadrons from colored partons
are scarce. A fundamental issue for hadronization models is the
conservation of 4-momentum as well as the entropy problem, because by
fusion/coalescence of massless (or low constituent mass) partons to
color neutral bound states of low invariant mass (e.g. pions) the
number of degrees of freedom and thus the total entropy is reduced in
the hadronization process. This problem - a
violation of the second law of thermodynamics as well as the
conservation of four-momentum and flavor currents - has been addressed
in Ref.~\cite{PRC08} on the basis of the DQPM employing covariant
transition rates for the fusion of 'massive' quarks and antiquarks to
color neutral hadronic resonances or strings. In fact, the dynamical
studies for an expanding partonic fireball in Ref.~\cite{PRC08}
suggest that the these problems have come to a practical solution.

A consistent dynamical approach - valid also for strongly interacting
systems - can be formulated on the basis of Kadanoff-Baym (KB)
equations~\cite{Sascha1} or off-shell transport equations in
phase-space representation,
respectively~\cite{Sascha1}. In the KB theory the field
quanta are described in terms of dressed propagators with complex
selfenergies.  Whereas the real part of the selfenergies can be
related to mean-field potentials (of Lorentz scalar, vector or tensor
type), the imaginary parts provide information about the lifetime
and/or reaction rates of time-like 'particles'~\cite{Crev}. Once the
proper (complex) selfenergies of the degrees of freedom are known the
time evolution of the system is fully governed by off-shell transport
equations (as described in Refs.~\cite{Sascha1,Crev}).  The
determination/extraction of complex selfenergies for the partonic
degrees of freedom has been performed before in
Ref.~\cite{Cassing07} by fitting lattice QCD (lQCD) 'data'
within the Dynamical QuasiParticle Model (DQPM). In fact, the DQPM
allows for a simple and transparent interpretation of lattice QCD
results for thermodynamic quantities as well as correlators and leads
to effective strongly interacting partonic quasiparticles with broad
spectral functions.  For a review on off-shell transport theory and
results from the DQPM in comparison to lQCD we refer the reader to
Ref.~\cite{Crev}.

The actual implementations in the PHSD transport approach have been
presented in detail in Refs.~\cite{PHSD,BCKL11}. Here we  present
results for transverse mass spectra and elliptic flow of hadrons for
heavy-ion collisions at relativistic energies in comparison to data
from the experimental collaborations.

\section{The PHSD approach}

The dynamics of partons, hadrons and strings in relativistic
nucleus-nucleus collisions is analyzed here within the
Parton-Hadron-String Dynamics approach~\cite{PRC08,PHSD,BCKL11}. In this
transport approach the partonic dynamics is based on Kadanoff-Baym
equations for Green functions with self-energies from the Dynamical
QuasiParticle Model (DQPM) \cite{Cassing07} which
describes QCD properties in terms of 'resummed' single-particle
Green functions. In Ref.~\cite{BCKL11}, the actual  three DQPM
parameters for the temperature-dependent effective coupling were
fitted to the recent lattice QCD results of Ref.~\cite{aori10}.
The latter lead to a critical temperature $T_c \approx$ 160 MeV
which corresponds to a critical energy density of $\epsilon_c
\approx$ 0.5 GeV/fm$^3$. In PHSD the parton spectral functions
$\rho_j$ ($j=q, {\bar q}, g$) are no longer $\delta-$ functions in
the invariant mass squared as in conventional cascade or transport
models but depend on the parton mass and width parameters:
\begin{equation}
 \rho_j(\omega,{\bf p})
 =
 \frac{\gamma_j}{E_j} \left(
   \frac{1}{(\omega-E_j)^2+\gamma_j^2} - \frac{1}{(\omega+E_j)^2+\gamma_j^2}
 \right)
\label{eq:rho}
\end{equation}
separately for quarks/antiquarks and gluons ($j=q,\bar{q},g$).
With the convention $E^2({\bf p}^2) = {\bf p}^2+M_j^2-\gamma_j^2$, the
parameters $M_j^2$ and $\gamma_j$ are directly related to the real
and imaginary parts of the retarded self-energy, {\it e.g.} $\Pi_j =
M_j^2-2i\gamma_j\omega$. The spectral function~(\ref{eq:rho}) is
antisymmetric in $\omega$ and normalized as
\begin{equation}
 \int_{-\infty}^{\infty} \frac{d \omega}{2 \pi} \
 \omega \ \rho_j(\omega, {\bf p}) = \int_0^{\infty} \frac{d
 \omega}{2 \pi} \ 2 \omega \ \rho_j(\omega, {\bf p}) = 1 \ .
\label{normalize}
\end{equation}

\begin{figure}[tbh]
\begin{minipage}[l]{ 7cm }
\phantom{a}\noindent
 \includegraphics[width=62mm]{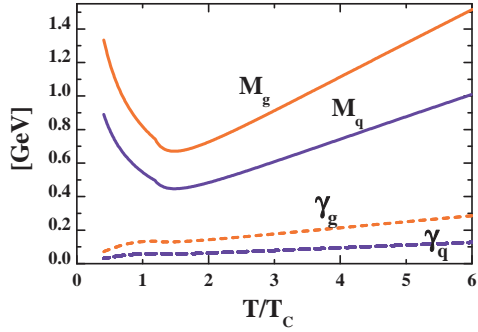}
\end{minipage}
\begin{minipage}[l]{ 4cm }
\caption{ The effective gluon mass
$M_g$ and witdh $\gamma_g$ as function of the scaled temperature $T/T_c$ (red lines).
The blue lines show the corresponding quantities for quarks.}
\label{fig1}
\end{minipage}
\end{figure}
\begin{figure}[thb]
\begin{minipage}[l]{ 7cm }
 \includegraphics[width=67mm]{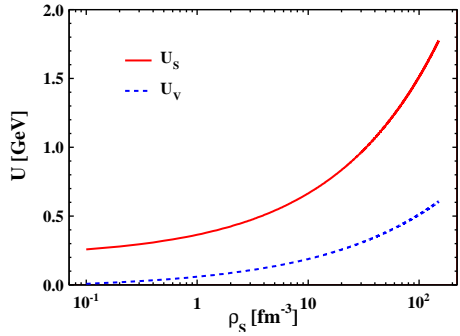}
\end{minipage}
\begin{minipage}[l]{ 4cm }
\caption{The scalar and vector mean-field potentials in the present
  PHSD model as a function of the scalar density $\rho_s$ of partons.}
\label{meanF}
\end{minipage}
\end{figure}

The actual parameters in Eq.~(\ref{eq:rho}), {\it i.e.} the gluon mass
$M_g$ and width $\gamma_g$ -- employed as input in the PHSD
calculations -- as well as the quark mass $M_q$ and width
$\gamma_q$, are depicted in Fig. \ref{fig1} as a
function of the scaled temperature $T/T_c$. As mentioned above these
values for the masses and widths have been fixed by fitting the
lattice QCD results from Ref.~\cite{aori10} in thermodynamic
equilibrium.

One might worry whether the quasiparticle properties - fixed in
thermal equilibrium - also should be appropriate for out-off
equilibrium configurations. This question is nontrivial and can only be answered
by detailed model investigations e.g. on the basis of Kadanoff-Baym
equations. We recall that such studies have been summarized in
Ref. \cite{Crev} for strongly interacting scalar fields that initially are
far off-equilibrium and simulate momentum distributions of colliding systems
at high relative momentum. The
results for the effective parameters $M$ and $\gamma$, which correspond to the time-dependent pole
mass and width of the propagator, indicate that the quasiparticle
properties - except for the very early off-equilibrium
configuration - are close to the equilibrium mass and width even
though the phase-space distribution of the particles is far from
equilibrium (cf. Figs. 8 to 10 in Ref. \cite{Crev}). Accordingly,
we will adopt the equilibrium quasiparticle properties also for
phase-space configurations out of equilibrium as  appearing in
relativistic heavy-ion collisions. The reader has to keep in mind
that this approximation is far from being arbitrary, however, not
fully equivalent to the exact solution.

We recall that the DQPM allows to extract a potential
energy density $V_p$ from the space-like part of the energy-momentum
tensor which can be tabulated {\it e.g.} as a function of the scalar
parton density $\rho_s$. Derivatives of $V_p$ with respect to $\rho_s$ then
define a scalar mean-field potential $U_s(\rho_s)$ which enters the
equation of motion for the dynamical partonic quasiparticles. As one
can see from Fig.~\ref{meanF}, the scalar potential is rather large
and nonlinearly increases with $\rho_s$. This implies that the
repulsive force due to $U_s(\rho_s)$ will change in a non-monotonous
way with the scalar density. The vector mean-field potential is not
negligible, too, especially at high $\rho_s$ and induces a Lorentz
force for the partons. Note that the vector mean-field vanishes with
decreasing scalar density whereas the scalar mean-field approaches a
constant value for $\rho_s\rightarrow$0.

Furthermore, a two-body
interaction strength can be extracted from the DQPM as well from the
quasiparticle width in line with Ref.~\cite{Andre}. The transition
from partonic to hadronic d.o.f. (and vice versa)  is
described by covariant transition rates for the fusion of
quark-antiquark pairs or three quarks (antiquarks), respectively,
obeying flavor current-conservation, color neutrality as well as
energy-momentum conservation~\cite{PHSD,BCKL11}. Since the dynamical
quarks and antiquarks become very massive close to the phase
transition, the formed resonant 'prehadronic' color-dipole states
($q\bar{q}$ or $qqq$) are of high invariant mass, too, and
sequentially decay to the groundstate meson and baryon octets
increasing the total entropy.

On the hadronic side PHSD includes explicitly the baryon octet and
decouplet, the $0^-$- and $1^-$-meson nonets as well as selected
higher resonances as in the Hadron-String-Dynamics (HSD)
approach~\cite{Ehehalt,HSD}. The color-neutral objects  
of higher masses ($>$1.5~GeV in case of baryonic states 
and $>$1.3~GeV in case of mesonic states) are treated as
`strings' (color-dipoles) that decay to the known (low-mass) hadrons
according to the JETSET algorithm~\cite{JETSET}. We discard an
explicit recapitulation of the string formation and decay and refer
the reader to the original work~\cite{JETSET}. Note that PHSD and HSD
(without explicit partonic degrees-of-freedom) merge at low energy
density, in particular below the critical energy density
$\varepsilon_c\approx$ 0.5~GeV/fm$^{3}$.

The PHSD approach was applied to nucleus-nucleus collisions from
$s_{NN}^{1/2} \sim$ 5 to 200 GeV in Refs.~\cite{PHSD,BCKL11} in order
to explore the space-time regions of 'partonic matter'. It was found
that even central collisions at the top-SPS energy of
$\sqrt{s_{NN}}=$17.3 GeV show a large fraction of nonpartonic, {\it
i.e.} hadronic or string-like matter, which can be viewed as a hadronic
corona~\cite{Aichelin}. This finding implies that neither hadronic nor
only partonic models can be employed to extract physical conclusions in
comparing model results with data.

\section{Application to nucleus-nucleus collisions}

\begin{figure}[t]
\phantom{a}\noindent
  \includegraphics[width=85mm]{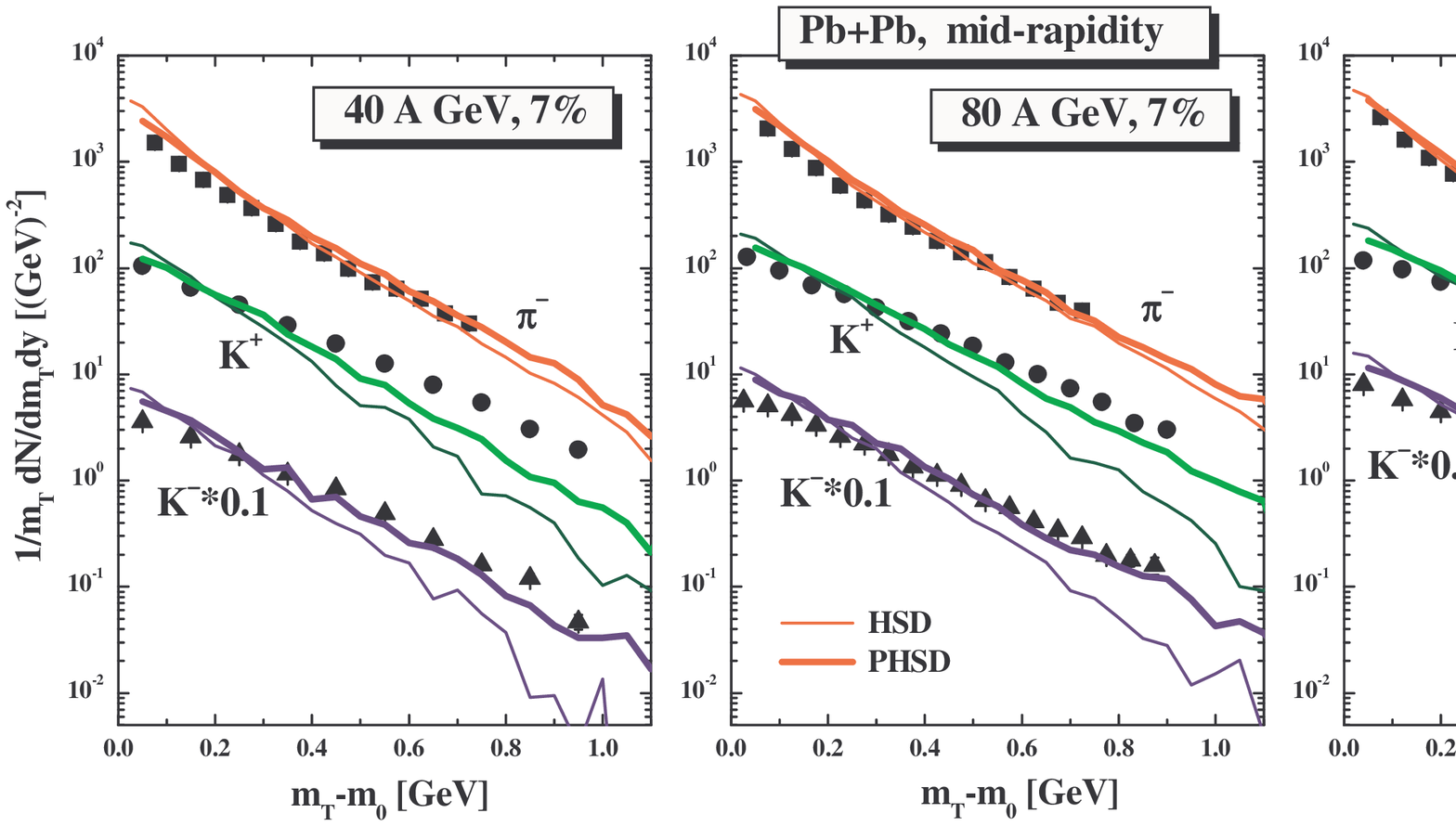}
  \caption{The $\pi^-$, $K^+$ and $K^-$ transverse mass spectra for
    central Pb+Pb collisions at 40, 80 and 158 A$\cdot$GeV from PHSD
    (thick solid lines) in comparison to the distributions from HSD
    (thin solid lines) and the experimental data from the NA49
    Collaboration~\cite{NA49a}. }
  \label{fig11}
\phantom{a}\vspace*{1mm}
\begin{minipage} [l] {6.5cm}
\includegraphics[width=50mm]{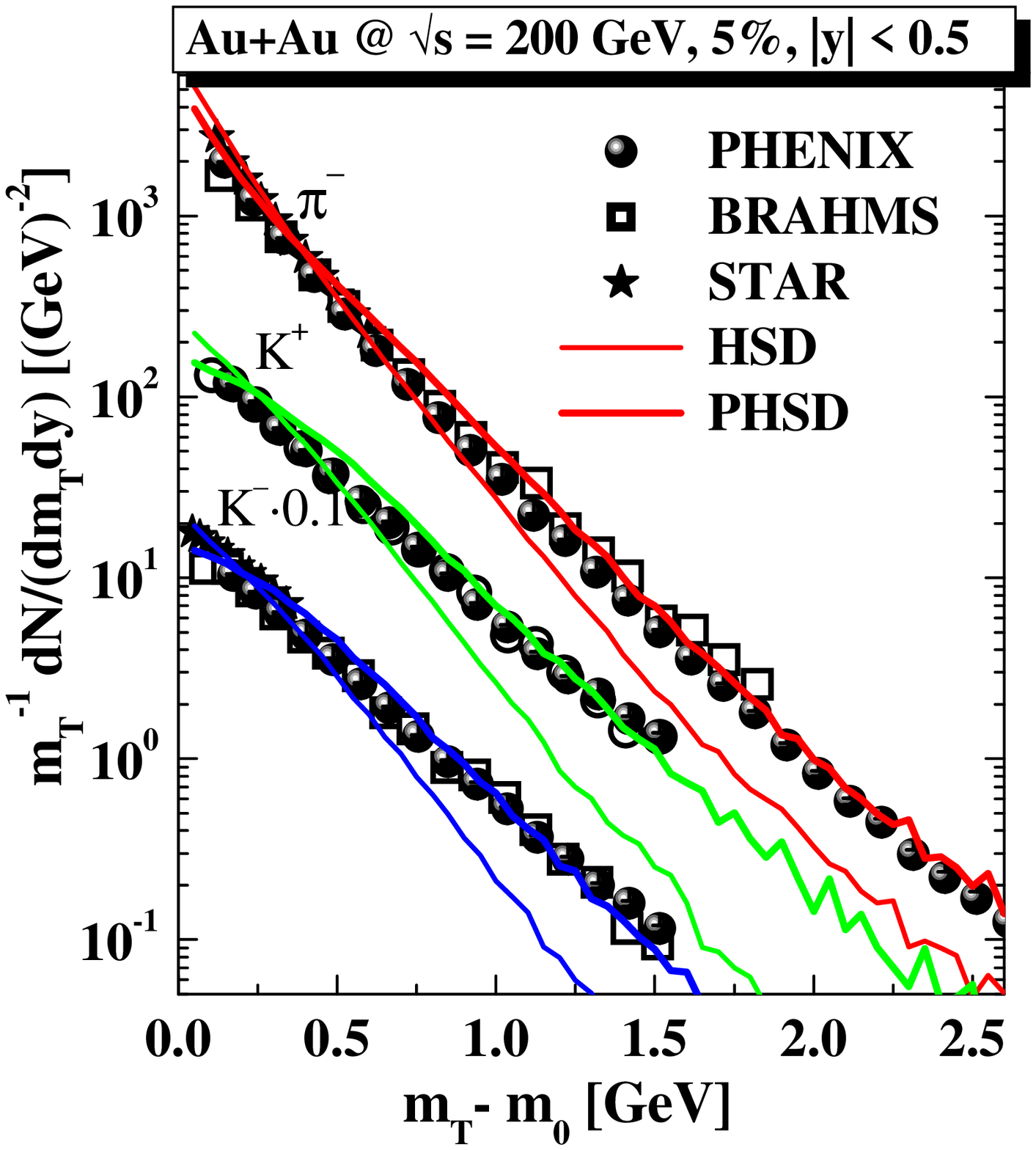}
\end{minipage}
\begin{minipage} [l] {4.5cm}
\caption{The $\pi^-$, $K^+$ and $K^-$ transverse mass spectra for 5\%
  central Au+Au collisions at $\sqrt{s}$ = 200 GeV from PHSD (thick solid
  lines) in comparison to the distributions from HSD (thin solid
  lines) and the experimental data from the BRAHMS, PHENIX and STAR
  Collaborations~\cite{PHENIX2,STAR3,BRAHMS} at midrapidity.}
\label{Fig_mtRHIC}
\end{minipage}
\end{figure}

In this Section we employ the PHSD approach to nucleus-nucleus
collisions at moderate relativistic energies.  It is of interest, how
the PHSD approach compares to the HSD~\cite{HSD} model (without
explicit partonic degrees-of-freedom) as well as to experimental
data. In Fig.~\ref{fig11} we show the transverse mass spectra of
$\pi^-$, $K^+$ and $K^-$ mesons for 7\% central Pb+Pb collisions at 40
and 80 A$\cdot$GeV and 5\% central collisions at 158 A$\cdot$GeV in
comparison to the data of the NA49 Collaboration~\cite{NA49a}.  Here
the slope of the $\pi^-$ spectra is only slightly enhanced in PHSD
relative to HSD which demonstrates that the pion transverse motion
shows no sizeable sensitivity to the partonic phase. However, the
$K^\pm$ transverse mass spectra are substantially hardened with
respect to the HSD calculations at all bombarding energies - i.e. PHSD
is more in line with the data - and thus suggests that partonic
effects are better visible in the strangeness-degrees of freedom.

The PHSD calculations for RHIC energies show a very similar trend -
the inverse slope increases by including the partonic phase -
cf. Fig. \ref{Fig_mtRHIC} where we show the transverse mass spectra of
$\pi^-$, $K^+$ and $K^-$ mesons for 5\% central Au+Au collisions at
$\sqrt{s}$ = 200 GeV in comparison to the data of the RHIC
Collaborations~\cite{PHENIX2,STAR3,BRAHMS}.

The hardening of the kaon spectra can be traced back to parton-parton
scattering as well as a larger collective acceleration of the partons
in the transverse direction due to the presence of repulsive vector
fields for the partons. The enhancement of the spectral slope for
kaons and antikaons in PHSD due to collective partonic flow shows up
much clearer for the kaons due to their significantly larger mass
(relative to pions). We recall that in Refs.~\cite{BratPRL} the
underestimation of the $K^\pm$ slope by HSD (and also UrQMD) had been
suggested to be a signature for missing partonic degrees of freedom;
the present PHSD calculations support this early suggestion.

The strange antibaryon sector is of further interest since here the HSD
calculations have always underestimated the yield~\cite{Geiss}.  Our
detailed studies in Ref.~\cite{PHSD} show that the HSD and PHSD
calculations both give a reasonable description of the $\Lambda +
\Sigma^0$ yield of the NA49 Collaboration~\cite{NA49_aL09}; both models
underestimate the NA57 data~\cite{NA57} by about 30\%. An even larger
discrepancy in the data from the NA49 and NA57 Collaborations is seen
for $(\bar \Lambda + \bar \Sigma^0)/N_{wound}$; here the PHSD
calculations give results which are in between the NA49 data and the
NA57 data whereas  HSD underestimates the $(\bar \Lambda + \bar
\Sigma^0)$ midrapidity yield at all centralities.

The latter result suggests that the partonic phase does not show up
explicitly in an enhanced production of strangeness (or in particular
strange mesons and baryons) but leads to a different redistribution of
antistrange quarks between mesons and antibaryons.  In fact, as
demonstrated in Ref.~\cite{PHSD}, we find no sizeable differences
in the double strange baryons from HSD and PHSD -- in a good agreement
with the NA49 data -- but observe a large enhancement in the double
strange antibaryons for PHSD relative to HSD.

 The anisotropy  in the azimuthal angle $\psi$ is usually characterized
by the even order Fourier coefficients $v_n =\langle exp(\, \imath \,
n(\psi-\Psi_{RP}))\rangle, \
 n = 2, 4, ...$, since for a smooth angular profile the odd harmonics
become equal to zero. As noted above, $\Psi_{RP}$ is the azimuth of the
reaction plane and the brackets denote averaging over particles
and events. In particular, for the widely used second order coefficient,
denoted as an elliptic flow, we have
 \be \label{eqv2}
 v_2 = \left<cos(2\psi-2\Psi_{RP})\right>=
 \left<\frac{p^2_x - p^2_y}{p^2_x + p^2_y}\right>~,
\ee
where $p_x$ and $p_y$ are the $x$ and $y$ components of the particle
momenta. This coefficient can be considered as a function of
centrality, pseudo-rapidity $\eta$ and/or transverse momentum $p_T$.  We
note that the reaction plane in PHSD is given by the $(x - z)$ plane
with the $z$-axis in the beam direction.

\begin{figure}[t]
\includegraphics[scale=.25]{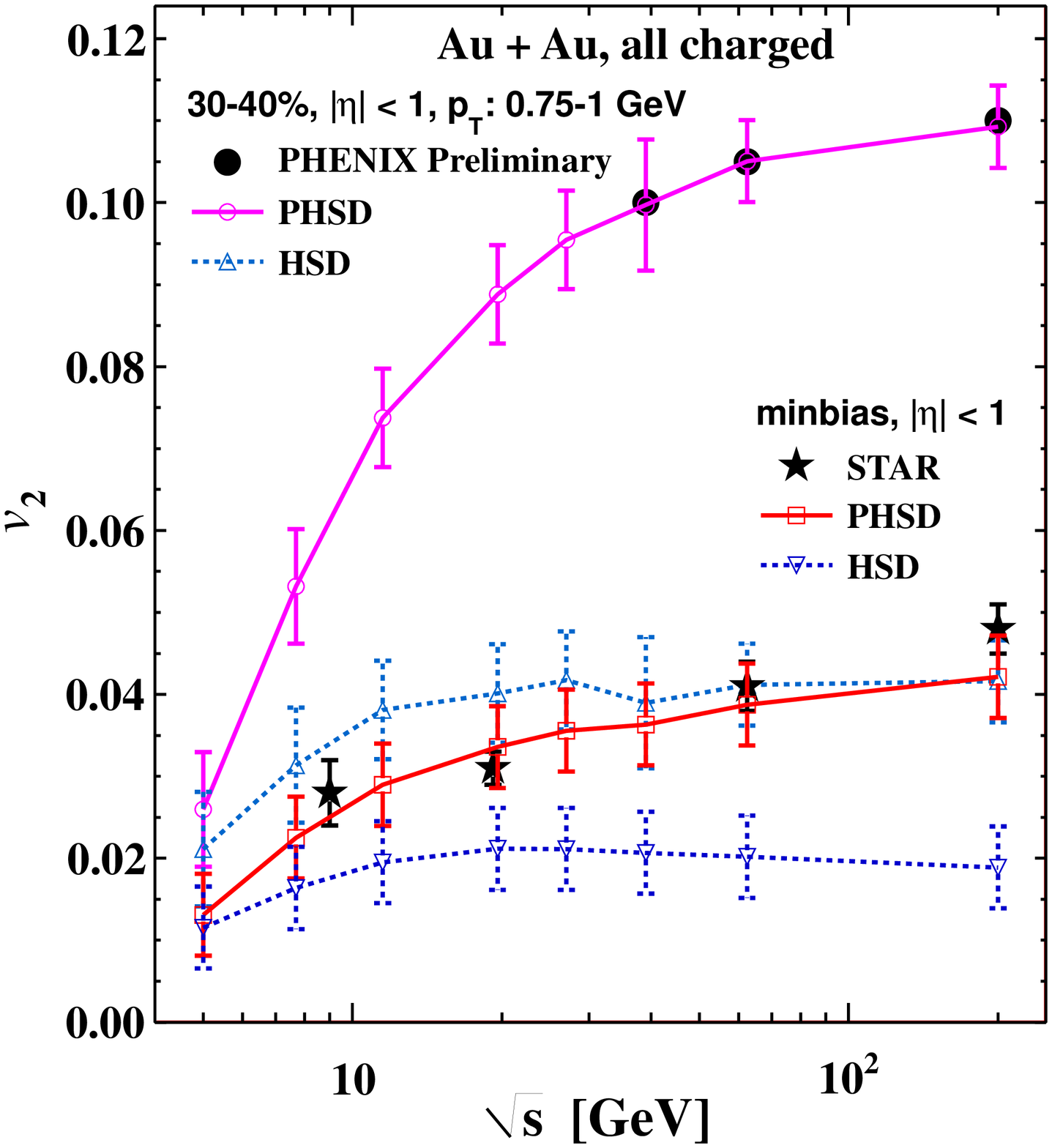}
\begin{minipage}{0.6\textwidth}
\phantom{a}\vspace*{-60mm}
\includegraphics[scale=.3]{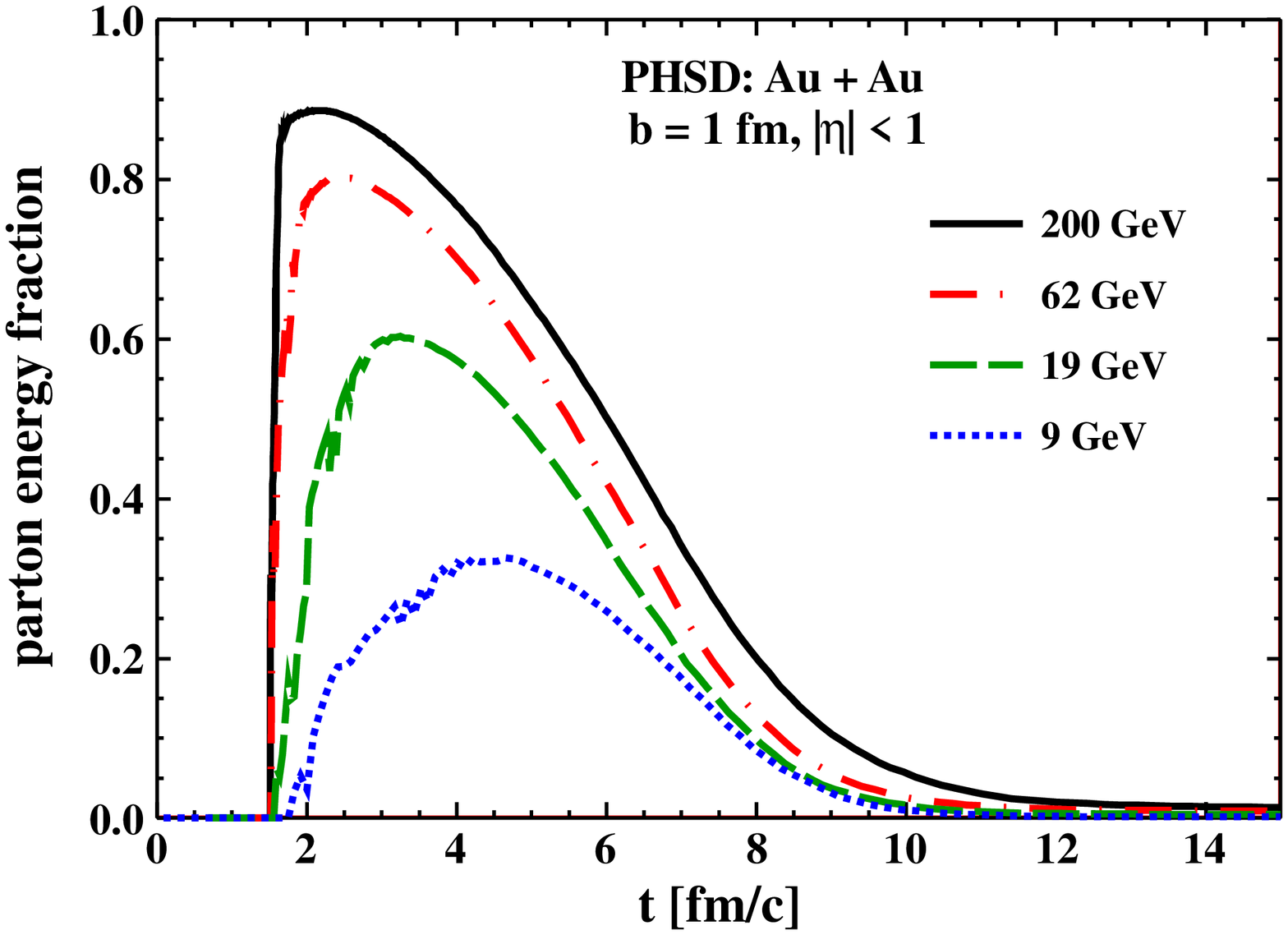}
\end{minipage}
\caption{(Left:)Average elliptic flow $v_2$ of charged particles at
midrapidity for two centrality selections calculated within the PHSD
(solid curves) and HSD (dashed curves). The $v_2$ STAR data
compilation for minimal bias collisions  are taken
from~\cite{NKKNM10} (stars) and the preliminary PHENIX
data~\cite{PHENIX_v2_s} are plotted by filled circles.
}
\label{vns}
\caption{(Right:) The evolution of the parton fraction of the total energy
  density at mid-pseudorapidity for different collision energies.}
\label{part}
\end{figure}

In Fig.~\ref{vns} the experimental $v_2$ excitation function in the
transient energy range is compared to the results from the PHSD
calculations \cite{Voka11}; HSD model results are given as well for
reference. We note that the centrality selection and acceptance are the
same for the data and models.

We recall that the  HSD model has been very successful in describing
heavy-ion spectra and rapidity distributions from SIS to SPS
energies. A detailed comparison of HSD results with respect to a
large experimental data set was  reported in
Refs.~\cite{mt-SIS,BratPRL,BRAT04}
for central Au+Au (Pb+Pb) collisions from SIS to top SPS energies.
Indeed, as shown in Fig.~\ref{vns} (dashed lines), HSD is in good
agreement with experiment for both data sets at the lower edge
($\sqrt{s_{NN}}\sim$10 GeV)  but predicts an approximately
energy-independent flow $v_2$ at  larger energies and, therefore, does
not match the experimental observations. This behavior is in quite
close agreement with another independent hadronic model, the UrQMD
(Ultra relativistic Quantum Molecular Dynamics)~\cite{UrQMD} (cf.
with Ref.~\cite{NKKNM10}).

\noindent

From the above comparison one may conclude  that the rise of $v_2$
with bombarding energy is not due to hadronic interactions and
models with partonic d.o.f. have to be addressed.
Indeed, the PHSD approach incorporates the parton medium effects in
line with a lQCD equation-of-state, as discussed above, and also
includes a dynamic hadronization scheme based on covariant
transition rates. It is seen from Fig.~\ref{vns} that PHSD performs
better: The elliptic flow $v_2$ from PHSD (solid curve) is fairly in
line with the data from the STAR and PHENIX collaborations and
clearly shows the growth of $v_2$ with the bombarding energy \cite{Voka11}.

The $v_2$ increase is clarified in Fig.~\ref{part} where the partonic
fraction of the energy density at mid-pseudorapidity with respect to
the total energy density in the same pseudorapidity interval is
shown. We recall that the repulsive scalar mean-field potential
$U_s(\rho_s)$ for partons in the PHSD model leads to an increase of
the flow $v_2$ as compared to that for HSD or PHSD calculations
without partonic mean fields.  As follows from Fig.~\ref{part}, the
energy fraction of the partons substantially grows with increasing
bombarding energy while the duration of the partonic phase is roughly
the same.

The $v_2$ coefficient measures the response of the heated and
compressed matter to the spatial deformation in the overlap region
of colliding nuclei, which is usually quantified by the
eccentricity $\epsilon_2=<y^2-x^2>/<x^2+y^2>$. Since the flow response
($v_2$) is proportional to the driving force ($\epsilon_2$), the
ratio $v_2/\epsilon_2$ is used to compare different impact
parameters and nuclei.

\begin{figure}[t]
\begin{minipage} [l] {7cm}
{\includegraphics[width=65mm]{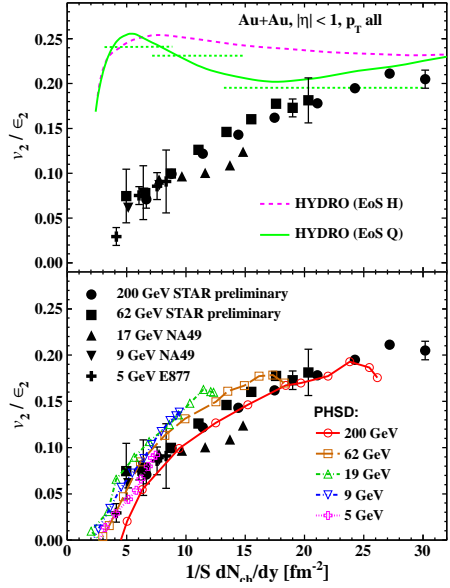}}
\end{minipage}
\begin{minipage} [l] {4.cm}
\caption{Scaling of $v_2/\epsilon_2$ vs $(1/S) (dN_{ch}/dy)$. The
  PHSD results \cite{Voka12} are given by lines with open symbols. Predictions of
  ideal boost-invariant hydrodynamics are shown in the top panel
  (from~\cite{Vol07}) and explained in the text. Our PHSD results are
  presented in the bottom panel. The experimental data points for
  Au+Au collisions at 200~GeV (circles) and 62~GeV (squares) are from
  Refs.~\cite{Vol07,Sh11}.}
\label{scal_v2}
\end{minipage}
\end{figure}

A remarkable property -- {\it universal scaling} -- has been proposed
in Ref.~\cite{VP00} (see Fig.~\ref{scal_v2}). It appears that
$v_2/\epsilon_2$ plotted versus $(1/S)dN_{ch}/dy$ falls on a
`universal' curve, which links very different regimes, ranging from
AGS to RHIC energies. Here
$S=\pi\sqrt{<x^2><y^2>}$ is the overlap area of the collision system
and $dN_{ch}/dy$ is the rapidity density of charged particles.

As seen from Fig. \ref{scal_v2} (lower pannel)  the universal scaling of
$v_2/\epsilon_2$ versus $(1/S) dN_{ch}/dy$  is approximately reproduced
by PHSD (see Ref. \cite{Voka12} for the details).  This feature is not
reproduced by hadronic transport models (such as HSD and UrQMD) and
meets (severe) problems in the various hydrodynamic descriptions as
demonstrated in the upper pannel of Fig. \ref{scal_v2} for a
pure hadronic equation of state ('EoS H') as well as with a QGP phase
transition ('EoS Q').

Thus, the experimentally observed scaling in Fig.~\ref{scal_v2} puts
very strong constraints on the initial microscopic
properties (entropy density, mean free path, {\it etc.}), as well as the
global longitudinal structure~\cite{Tor07}.

\vspace*{3mm}
Work supported in part by the HIC for FAIR framework of the LOEWE
program and by DFG.

\end{document}